\documentstyle[12pt,epsfig]{article}
\def\lbldef#1#2{\expandafter\gdef\csname #1\endcsname {#2}}

\def\href#1#2{#2}  

\begin{document}
\baselineskip=15.5pt
\pagestyle{plain}
\setcounter{page}{1}

\begin{titlepage}

\begin{flushright}
CERN-TH/99-409\\
hep-th/9912175
\end{flushright}
\vspace{10 mm}

\begin{center}
{\Large Probing Solitons in Brane Worlds}

\vspace{5mm}

\end{center}

\vspace{5 mm}

\begin{center}
{\large Donam Youm\footnote{E-mail: Donam.Youm@cern.ch}}

\vspace{3mm}

Theory Division, CERN, CH-1211, Geneva 23, Switzerland

\end{center}

\vspace{1cm}

\begin{center}
{\large Abstract}
\end{center}

\noindent

We study dynamics of a probe $p$-brane and a test particle in 
the field background of fully localized solutions describing 
the source $p$-brane within the worldvolume of the source 
domain wall.  We find that the probe dynamics in the background 
of the source $p$-brane in one lower dimensions is not reproduced, 
indicating that $p$-branes within the worldvolume of domain walls 
perhaps describe an exotic phase of $p$-branes in brane worlds.  We 
speculate therefore that a $(p+1)$-brane where one of its longitudinal 
directions is along the direction transverse to the domain wall is 
the right description of the $p$-brane in the brane world with the  
expected properties.

\vspace{1cm}
\begin{flushleft}
CERN-TH/99-409\\
December, 1999
\end{flushleft}
\end{titlepage}
\newpage

\section{Introduction}

Recently, a lot of attention has been paid to the idea on non-compact 
compactification \cite{ncm1,ncmn,ncm2,ncm3,ncm4,ncm5,ncm6,ncm7} since it was 
observed by Randall and Sundrum (RS) \cite{rs1,rs2,rs3} that such old 
idea modified by the modern brane language can be applied to solve the 
hierarchy problem in particle physics.  According to the RS model, our 
four-dimensional world is regarded as being confined within a 
(non-dilatonic) domain wall.  In the RS model, the graviton Kaluza-Klein 
(KK) spectrum consists of a normalizable zero mode bound state 
(identified as the four-dimensional massless graviton) and a continuum 
of massive KK modes.  Because the continuum of massive KK modes is 
extremely suppressed, Newton's $1/r^2$ law of four-dimensional gravity 
is reproduces with the correction from the massive KK modes being beyond 
the limit of current experimental precision.  

In Ref. \cite{youm}, it is found out that the RS model can be extended 
(with the exactly same structure of graviton KK spectrum) to the dilatonic 
domain wall case if the dilaton coupling parameter is sufficiently 
small.  Dilatonic generalization (where the cosmological constant 
term in the action is multiplied by the dilaton factor) is a natural 
generalization of the RS model and is particularly interesting, because 
most of the massive and the gauged supergravity theories contain 
dilaton factor in the cosmological constant term.  (Few of gauged 
supergravity theories that have cosmological constant terms without 
dilaton factor and therefore admit the AdS space as a solution  
are those obtained by compactifying the eleven-dimensional 
supergravity on $S^4$ or $S^7$ and the type-IIB supergravity on 
$S^5$.)  It has been observed \cite{sg1,sg2,sg3,sg4} that all the 
supersymmetric domain wall solutions that have so far been constructed  
within the five-dimensional gauged supergravity with non-dilatonic 
cosmological constant term have the undesirable exponentially increasing 
warp factor, rather than the exponentially decreasing warp factor (which 
enables nice trapping of gravity within the domain wall) of the RS model.  
Perhaps, the dilatonic generalization of the RS type model will provide 
with the supersymmetric embedding into the gauged supergravity theories.  
The dilatonic generalization also seems to be more desirable, because 
the most of the domain wall solutions obtained by compactifying 
(intersecting) branes in string theories are dilatonic.  It is one of 
the purposes of this paper to elaborate more on the most of possible 
dilatonic domain walls in string theories and gauged supergravity theories 
in relation to the RS type model.

Since it is observed \cite{rs2} that four-dimensional gravity is 
recovered within the domain wall of the RS model, it is of interest to 
study gravitating objects in domain walls to see whether 
lower-dimensional physics (in the worldvolume the domain walls) is 
reproduced.  (Previous related works are, for example, Refs. 
\cite{hwa,hm1,gs,hm2}.)  For this purpose, fully localized solutions 
describing extreme branes within the worldvolume of extreme domain walls are 
constructed in Ref. \cite{youm}.  It is a purpose of this paper to study 
dynamics of a probe $p$-brane and an uncharged test particle in 
such background.  Unfortunately, the fields produced by such configurations 
have properties different from those produced by the corresponding branes 
in one lower dimensions.  So, it seems that such localized solutions are 
not suitable for describing branes in lower-dimensional world with the 
right expected properties.  On the other hand, existence of fully localized 
solutions where branes live within the worldvolume of domain walls seems to 
indicate that (if the RS model is a correct description of nature) such 
solutions may describe exotic phase of branes.  We propose that the 
higher-dimensional origin of $p$-branes in one lower dimensions (with 
the right expected properties in one lower dimensions) should rather be 
identified as $(p+1)$-branes where one of their longitudinal directions 
is along the direction transverse to the domain wall.  

The paper is organized as follows.  In section 2, we summarize fully 
localized solution constructed in Ref. \cite{youm} and elaborate on 
possible dilatonic domain wall solutions in massive and gauged supergravity 
theories.  In section 3, we study dynamics of a probe $p$-brane moving in 
the field background of the source $p$-brane living in the worldvolume 
of the $D$-dimensional source domain wall and compare it to the dynamics in 
the background of the $(D-1)$-dimensional source $p$-brane.  In section 4, 
we repeat the similar analysis with a test particle moving in the same 
backgrounds.  The conclusion is given in section 5.

\section{Dilatonic Domain Walls in Supergravity Theories}

In this section, we discuss various extreme dilatonic domain wall 
solutions that occur in supergravity and string theories and 
localized solutions describing extreme $p$-brane within extreme 
dilatonic domain walls.   

We begin by considering the following $D$-dimensional Einstein-frame 
action for the system of the graviton $g^E_{\mu\nu}$ ($\mu,\nu=0,1,...,D-1$), 
the dilaton $\phi$, the $(p+1)$-form potential with the field strength 
$F_{p+2}=dA_{p+1}$ and the $(D-1)$-form potential $A_{D-1}$ with the field 
strength $F_D=dA_{D-1}$:
\begin{equation}
S_E={1\over{2\kappa^2_D}}\int d^Dx\sqrt{-g^E}\left[{\cal R}_{g^E}-{4\over{D-2}}
(\partial\phi)^2-{1\over{2\cdot (p+2)!}}e^{2a_p\phi}F^2_{p+2}
-{1\over{2\cdot D!}}e^{2a\phi}F^2_D\right],
\label{einpbrnindw}
\end{equation}
where $a_p$ and $a$ are respectively the dilaton coupling parameters for the 
$(p+1)$-form and the $(D-1)$-form potentials.  Through a Poincar\'e 
dualization, one can replace $F_D$ with the cosmological constant $\Lambda$, 
thereby the kinetic term for $A_{D-1}$ in the above action is replaced by 
the cosmological constant term $e^{-2a\phi}\Lambda$.  

The solution to the field equations of the action (\ref{einpbrnindw}) 
describing the extreme dilatonic $p$-brane with the longitudinal coordinates 
${\bf w}=(w_1,...,w_p)$ within the worldvolume of the extreme dilatonic 
domain wall with the longitudinal coordinates $({\bf w},{\bf x})$, where 
${\bf x}=(x_1,...,x_{D-p-2})$, has the following form:
\begin{eqnarray}
ds^2_E&=&H^{4\over{(D-2)\Delta}}\left[H^{-{{4(D-p-3)}\over{(D-2)
\Delta_p}}}_p\left(-dt^2+dw^2_1+\cdots+dw^2_p\right)\right.
\cr
& &\left.+H^{{4(p+1)}\over{(D-2)\Delta_p}}_p\left(dx^2_1+\cdots+dx^2_{D-p-2}
\right)\right]+H^{{4(D-1)}\over{(D-2)\Delta}}H^{{4(p+1)}\over
{(D-2)\Delta_p}}_pdy^2,
\cr
e^{2\phi}&=&H^{{(D-2)a}\over{\Delta}}H^{{(D-2)a_p}\over{\Delta_p}}_p, 
\cr
A_{tw_1...w_p}&=&{2\over\sqrt{\Delta_p}}\left(1-H^{-1}_p\right),
\ \ \ \ \ \ 
A_{tw_1...w_px_1...x_{D-p-2}}={2\over\sqrt{\Delta}}\left(1-H^{-1}\right).
\label{pbrdwsol}
\end{eqnarray}
The harmonic functions $H_p$ and $H$ for the $p$-brane and the domain wall 
satisfy the following coupled partial differential equations:
\begin{equation}
\partial^2_yH_p+H\partial^2_{\bf x}H_p=0,\ \ \ \ \ \ \ 
\partial^2_yH=0,
\label{cpdiffeq}
\end{equation}
and therefore are given by \cite{lsol,youm}:
\begin{equation}
H_p=1+{{Q_p}\over{\left[|{\bf x}|^2+{4\over{9Q^2}}(1+Q|y|)^3\right]^{{3(D-p)-8}
\over 6}}},\ \ \ \ \ \ \ \ \  H=1+Q|y|,
\label{harmfncs2}
\end{equation}
where $Q$ in the harmonic function $H$ is related to the cosmological 
constant term $\Lambda$ as $\Lambda=-2Q^2/\Delta$.

Note, although derived from the field equations of the action of the specific 
form (\ref{einpbrnindw}) with only metric, dilaton and form fields, the 
solution (\ref{pbrdwsol}) with the harmonic functions (\ref{harmfncs2}) 
generically describe any extreme single-charged $p$-brane within the 
worldvolume of single-charged extreme domain wall in string theories, which 
has additional scalar fields originated from the internal components 
of metric and form potentials.  

First of all, when viewed as a solution to the field equations of the action 
(\ref{einpbrnindw}), the parameters $\Delta_p$ and $\Delta$ of the solution 
(\ref{pbrdwsol}) take the following forms determined by the dilaton coupling 
parameter $a_p$ or $a$, as well as by $p$ and $D$:
\begin{eqnarray}
\Delta_p&=&{{(D-2)a^2_p}\over 2}+{{2(p+1)(D-p-3)}\over{D-2}},
\cr
\Delta&=&{{(D-2)a^2}\over{2}}-{{2(D-1)}\over{D-2}},
\label{deltavals}
\end{eqnarray}
and the consistency of the field equations \cite{ir} or the no-force  
requirement \cite{nf} restricts the dilaton coupling parameters to satisfy 
the following constraint:
\begin{equation}
aa_p={{4(p+1)}\over{(D-2)^2}}.
\label{bpara}
\end{equation}

In general, for $p$-brane solutions in string theories in $D<10$, 
there are additional scalars associated with the internal components 
of the spacetime metric and form potentials, in addition to the 
$D$-dimensional string theory dilaton (defined in terms of the dilaton 
in ten dimensions and the determinant of the internal part of the 
metric), and the kinetic terms for the form potentials are multiplied not 
only by the dilaton (with an appropriate dilaton coupling parameter) but 
also by other scalars in some cases.  Nevertheless, one can view the 
solution (\ref{pbrdwsol}) as the metric $g^E_{\mu\nu}$, $D$-dimensional 
string theory dilaton $\phi$ and form fields $A_{p+1}$ and $A_{D-1}$ part 
of full supergravity solution describing the BPS $p$-brane within the BPS 
domain wall in a $D$-dimensional string theory
\footnote{Of course, one can linearly combine scalar fields to bring 
the action to the form (\ref{einpbrnindw}) with a single dilatonic scalar, 
but for the convenience of identifying string theory dilaton in the 
solutions we shall just regard the solution (\ref{pbrdwsol}) as a 
part of full string theory solution, with $\phi$ being regarded as the 
$D$-dimensional string theory dilaton, rather than as a dilatonic scalar 
defined as a linear combination of all the scalars.}.  
In this case, the parameters $\Delta_p$ and $\Delta$ in the solution 
(\ref{pbrdwsol}) are no longer restricted to take the forms 
(\ref{deltavals}) determined by the dilaton coupling parameters, $p$ and 
$D$, and the dilaton coupling parameters $a_p$ and $a$ are not constrained 
by the relation (\ref{bpara}).  However, they take the forms determined by 
the type of charges that the solution carries and the number $N$ of 
constituent branes, which have the same magnitude of charges.  

To see this easily, we apply the Weyl-scaling transformation $g^E_{\mu\nu}=
e^{-{4\over{D-2}}\phi}g_{\mu\nu}$ to the solution (\ref{pbrdwsol}).  
In this new frame, the Einstein term and the dilaton kinetic term of the 
Einstein-frame action (\ref{einpbrnindw}) take the following string-frame 
form for a $D$-dimensional effective string theory action:
\begin{equation}
{1\over{2\kappa^2_D}}\int d^D\sqrt{-g}e^{-2\phi}\left[{\cal R}_g+
4\partial_{\mu}\phi\partial^{\mu}\phi\right].
\label{eindilterm}
\end{equation}
This is the main reason why we choose to normalize the dilaton $\phi$ to 
have the coefficient $4/(D-2)$ in the dilaton kinetic term in the 
Einstein-frame action (\ref{einpbrnindw}) instead of the usual 
canonical choice of $1/2$.  In this ``string-frame'', the spacetime metric 
for the extreme $p$-brane solution takes the following form:
\begin{eqnarray}
ds^2&=&H^{-{{4(D-p-3)-2(D-2)a_p}\over{(D-2)\Delta_p}}}_p\left[-dt^2+dx^2_1+
\cdots+dx^2_p\right]
\cr
& &\ \ \ +H^{{4(p+1)+2(D-2)a_p}\over{(D-2)\Delta_p}}_p
\left[dy^2_1+\cdots+dy^2_{D-p-1}\right].
\label{strpbrnmet}
\end{eqnarray}
By noting that in the KK compactification of string theories on any 
Ricci-flat manifolds with the KK gauge fields set equal to zero the 
$D$-dimensional string-frame metric $g_{\mu\nu}$ ($\mu,\nu=0,1,...,D-1$) 
is related to the ten-dimensional string-frame metric $G_{MN}$ ($M,N=
0,1,...,9$) as $g_{\mu\nu}=G_{\mu\nu}$, one can find generic expressions 
for $a_p$ and $\Delta_p$ for any $p$-brane solutions in string 
theory on any Ricci-flat compactification manifolds.  Namely, one can 
determine $a_p$ and $\Delta_p$ for various cases by comparing the 
metric (\ref{strpbrnmet}) to the $D$-dimensional uncompactified part 
of the ten-dimensional string-frame metric for (intersecting) BPS 
branes (with equal magnitudes of constituent branes).  
The resulting general rules for determining values of the parameters $a_p$ 
and $\Delta_p$ in the $p$-brane solutions in $D$-dimensional string 
theories on any Ricci-flat manifolds are as follows:
\begin{itemize}
\item For a $p$-brane made out of $N$ constituent branes, $\Delta_p=4/N$.

\item For a $p$-brane made out of $N$ constituent branes of the same type of 
charges with the same magnitude, the values of the dilaton coupling parameter 
$a_p$ are as follows:
\begin{displaymath}
\begin{array}{ll}
a_p={{D-2p-4}\over{D-2}}, & \textnormal{originated from intersecting 
D-branes}\\
a_p={{2(D-p-3)}\over{D-2}}, & \textnormal{originated from intersecting 
NS5-branes}\\
a_p=-{{2(p+1)}\over{D-2}}, & \textnormal{originated from intersecting 
fundamental strings}
\end{array}
\end{displaymath}

\item For a $D$-dimensional $p$-brane made out of $N_i$ numbers of type $i$ 
branes, where $i$ stands for D-brane, NS5-brane or the fundamental string, 
with the same magnitude of charges, the dilaton coupling parameter $a_p$ is 
the weighted average of the dilaton coupling parameters $a^i_p$ of each 
type of constituent, namely,
\begin{equation}
a_p={{\sum_iN_ia^i_p}\over{\sum_iN_i}},
\label{gendilpara}
\end{equation}
where the expression for $a^i_p$ for each case is given in the previous item. 
\end{itemize}

The domain wall solutions in string theories can be constructed 
\cite{dw,pop1,pop2,pop3} through the ordinary compactification of 
(intersecting) branes (with the same magnitudes of charges for the constituent 
branes) on a Ricci flat manifold combined with the Scherk-Schwarz 
compactification \cite{ss}, which leads to the massive supergravity theories 
with the cosmological constant term.  For such compactification, the 
parameters $\Delta$ and $a$ in the domain wall solution take the forms 
corresponding to the $p=D-2$ case of the BPS $p$-brane solutions discussed 
in the previous paragraph.  Namely, the ordinary compactification on a 
Ricci-flat manifold combined with the Scherk-Schwarz compactification leads 
to the domain wall solutions with $\Delta=4/N>0$ ($N\in{\bf Z}^+$), only.  
On the other hand, the domain wall solutions of the massive supergravity 
theories which cannot be obtained through the Scherk-Schwarz type 
compactifications of the eleven- or ten-dimensional supergravity theory 
have $\Delta\neq 4/N$. (Generally, such domain wall solutions 
typically have $\Delta<0$.)  Some of such massive supergravity theories 
in $D<10$ can be constructed through the Freund-Rubin compactification 
\cite{fr} of supergravity theories in $D=10,11$ on spheres.  (It is also 
pointed out in Ref. \cite{youm} that the combined Scherk-Schwarz type 
and spherical compactification
\footnote{In the spherical compactifications of branes, one goes to the 
near horizon region of the (intersecting) branes and compactifies all the 
angular coordinates of the (overall) transverse space in the spherical 
coordinates on a sphere.  So, (intersecting) branes can be interpreted 
as interpolating between the Minkowski vacuum (at infinity) and the 
vacuum of gauged supergravity theory (in the near horizon region) 
\cite{gt,ct}.} 
of (intersecting) branes (with the equal magnitudes of the constituent 
brane charges) leads to dilatonic domain wall solutions with $\Delta\neq 
4/N$, constructed in Refs. \cite{lps,lpt,town,bbh}.)  
Examples are gauged supergravities in $D=7$ \cite{pn}, $D=6$ \cite{at,rom} 
and $D=5$ \cite{gst}, and the $SU(2)\times SU(2)$ gauged supergravity in 
$D=4$ \cite{fs}, whose domain wall solutions all have $\Delta=-2$.  For 
example, the domain wall solutions of the gauged supergravity in $D=7$ 
\cite{pn} [of the $SU(2)\times SU(2)$ gauged supergravity in $D=4$ \cite{fs}] 
can be obtained by starting from the intersecting two NS5-branes on a line 
and then compactifying on $S^3$ [on $S^3\times S^3$] in the near horizon 
region of one of the NS5-branes and far away from the other NS5-brane [in 
the near horizon of both of the NS5-branes] \cite{ct}.  
These exceptional cases are particularly interesting, because domain wall 
solutions in gauged supergravity theories (sometimes constructible 
through the spherical compactifications of branes in string theories) 
provide with dilatonic generalization \cite{youm} of the RS model 
\cite{rs1,rs2,rs3}.  Although it is found out \cite{youm} that dilatonic 
domain walls with $\Delta<-2$ have the KK graviton spectrum similar 
to that of the RS domain wall solution, thereby leading to the 
generalization of the RS model to the dilatonic domain walls with $\Delta<-2$, 
it may turn out that even in the $\Delta<0$ case the boundary condition on 
the wave function at the location of the domain wall (due to the 
$\delta$-function potential in the Schr\"odinger equation) excludes 
undesirable non-zero mode KK states to reproduce lower-dimensional gravity 
with extremely suppressed contribution from the non-zero graviton KK modes.  
To sum up, in order to construct domain wall solutions in string theories 
which generalize the RS model and supergravity solutions describing branes 
in such domain walls, one starts with (intersecting) branes in string 
theories and applies the Freund-Rubin compactification on spheres, along 
with the ordinary KK compactification when necessary.

\section{Dynamics of a Probe $p$-Brane}

In this section, we study dynamics of a probe $p$-brane moving in the 
background (\ref{pbrdwsol}) of the source $p$-brane living in the brane 
world, comparing with the dynamics of a probe $p$-brane moving in 
the following $(D-1)$-dimensional source $p$-brane background:
\begin{eqnarray}
ds^2_E&=&H^{-{{4(D-p-4)}\over{(D-3)\Delta_p}}}_p\left[-dt^2+dw^2_1+\cdots+
dw^2_p\right]+H^{{4(p+1)}\over{(D-3)\Delta_p}}_p\left[dx^2_1+\cdots+
dx^2_{D-p-2}\right],
\cr
e^{2\phi}&=&H^{{(D-3)a_p}\over{\Delta_p}}_p,\ \ \ \ \ \ 
A_{tx_1...x_p}={2\over\sqrt{\Delta_p}}\left(1-H^{-1}_p\right),
\label{pbrnsol}
\end{eqnarray}
where
\begin{equation}
H_p=1+{{Q_p}\over{|{\bf x}|^{D-p-4}}},\ \ \ \ 
\Delta_p={{(D-3)a^2_p}\over 2}+{{2(p+1)(D-p-4)}\over{D-3}}.
\label{harmdeldefs}
\end{equation}

The worldvolume action for a dilatonic $p$-brane with the following bulk 
action:
\begin{equation}
S_E={1\over{2\kappa^2_D}}\int d^Dx\sqrt{-g^E}\left[{\cal R}_{g^E}-{4\over{D-2}}
(\partial\phi)^2-{1\over{2\cdot (p+2)!}}e^{2a_p\phi}F^2_{p+2}\right]
\label{dilpbrnact}
\end{equation}
has the following form:
\begin{eqnarray}
S_{\sigma}&=&-T_p\int d^{p+1}\xi\left[e^{-a_p\phi}\sqrt{-{\rm det}\,
\partial_aX^{\mu}\partial_bX^{\nu}g^E_{\mu\nu}}\right.
\cr
& &\ \ \ \ \ \ \ \ \ \ \ \ 
\left.+{\sqrt{\Delta_p}\over 2}{1\over{(p+1)!}}\epsilon^{a_1\dots a_{p+1}}
\partial_{a_1}X^{\mu_1}\dots\partial_{a_{p+1}}X^{\mu_{p+1}}
A_{\mu_1\dots\mu_{p+1}}\right],
\label{wvdpbrnact}
\end{eqnarray}
where the target space fields $g^E_{\mu\nu}$, $\phi$ and $A_{\mu_1\dots
\mu_{p+1}}$ are the background fields (produced by the source brane) in 
which the probe $p$-brane with the target space coordinates $X^{\mu}$ 
($\mu=0,1,...,D-1$) and the worldvolume coordinates $\xi^a$ ($a=0,1,...,p$) 
moves.  Note, the scalar $\phi$ in the above is not the string theory 
dilaton, but is a linear combination of all the nontrivial scalars 
of the solution.

In the static gauge, in which $X^a=\xi^a$, the pull-back fields for the 
probe $p$-brane, oriented in the same way as the source $p$-brane, 
take the following forms:
\begin{eqnarray}
\hat{G}_{ab}&\equiv&g^E_{\mu\nu}\partial_aX^{\mu}\partial_bX^{\nu}=
g^E_{ab}+g^E_{ij}\partial_aX^i\partial_bX^j,
\cr
\hat{A}_{a_1...a_{p+1}}&\equiv&A_{\mu_1...\mu_{p+1}}\partial_{a_1}X^{\mu_1}
...\partial_{a_{p+1}}X^{\mu_{p+1}}=A_{a_1...a_{p+1}},
\label{pullbckst}
\end{eqnarray}
where the indices $i,j=1,...,D-p-1$ label the transverse space of the 
probe $p$-brane, i.e., $(X^i)=(x_1,...,x_{D-p-2},y)$ in the notation of Eq. 
(\ref{pbrdwsol}).  So, the worldvolume action (\ref{wvdpbrnact}) takes the 
following form:
\begin{equation}
S_{\sigma}=-T_p\int d^{p+1}\xi\left[e^{-a_p\phi}\sqrt{-{\rm det}
\left(g^E_{ab}+g^E_{ij}\partial_aX^i\partial_bX^j\right)}+{\sqrt{\Delta_p}
\over 2}A_{01...p}\right].
\label{wvsimpact}
\end{equation}
From now on, we assume that the target space transverse coordinates 
$X^i$ for the probe $p$-brane depend on the time coordinate $\tau=\xi^0$ 
only, i.e., $X^i=X^i(\tau)$.

In the target space background (\ref{pbrdwsol}) of the extreme $p$-brane 
localized within the domain wall, the probe $p$-brane action (\ref{wvsimpact}) 
takes the following form:
\begin{equation}
S_{\sigma}=-T_p\int d^{p+1}\xi\left[H^{-1}_p\sqrt{1-H^{4\over{\Delta_p}}_p
v^2_{\parallel}-H^{4\over\Delta}H^{4\over{\Delta_p}}_pv^2_{\perp}}
+1-H^{-1}_p\right],
\label{spact}
\end{equation}
where $v_{\parallel}$ and $v_{\perp}$ are respectively the velocities of the 
probe $p$-brane in the longitudinal ${\bf x}$ and the transverse $y$ 
directions of the domain wall:
\begin{equation}
|v_{\parallel}|\equiv\sqrt{\sum^p_{i=1}\left({{dx_i}\over{d\tau}}\right)^2},
\ \ \ \ \ \ \ 
v_{\perp}\equiv{{dy}\over{d\tau}},
\label{spddefs}
\end{equation}
and we used Eqs. (\ref{deltavals}) and (\ref{bpara}), which hold for the 
dilatonic brane solution (\ref{pbrdwsol}) to the field equations of the 
action (\ref{einpbrnindw}), to simplify the expression.  For the 
source-probe method \cite{nf,dps,lt,kir1,kir2,kir3} to be valid, one has to 
assume that the source brane is much heavier than the probe brane so that the 
backreaction of the probe brane to the source brane can be negligible.  So, 
the constant term in the harmonic function for the source $p$-brane can 
be neglected.  One further assumes that the velocity of the probe 
$p$-brane is very small ($v_{\parallel}\approx 0\approx v_{\perp}$) 
and changes very slowly so that the radiation will be negligible, 
allowing quasistatic evolution of the system described by the geodesic 
motion in the moduli space.  In this limit, the probe action (\ref{spact}) 
is approximated to
\begin{equation}
S_{\sigma}=-m_p\int d\tau+{{m_p}\over 2}\int d\tau
\left[H^{{4-\Delta_p}\over{\Delta_p}}_pv^2_{\parallel}+
H^{4\over\Delta}H^{{4-\Delta_p}\over{\Delta_p}}_pv^2_{\perp}\right]
+{\cal O}(v^4),
\label{smllvelact}
\end{equation}
where $m_p$ is the mass of the probe $p$-brane given by the product of 
the probe $p$-brane tension $T_p$ and the volume factor resulting 
from the integration with respect to $\xi^a$ ($a=1,...,p$).  
So, one can see that the motion of the probe $p$-brane in the background 
(\ref{pbrdwsol}) is given by the geodesic motion in the moduli space with 
the following moduli metric:
\begin{equation}
ds^2_p=H^{{4-\Delta_p}\over{\Delta_p}}_pd{\bf x}\cdot d{\bf x}+
H^{4\over\Delta}H^{{4-\Delta_p}\over{\Delta_p}}_p dz^2.
\label{modmet}
\end{equation}
The probe action and the moduli metric for a probe $p$-brane moving 
in the background (\ref{pbrnsol}) of the $(D-1)$-dimensional source 
$p$-brane are also given by Eqs. (\ref{smllvelact}) and (\ref{modmet}) 
with $H=1$ and $H_p$ given in Eq. (\ref{harmdeldefs}). 

For the motion of the probe in the ${\bf x}$-direction, we set all the angular 
momenta of the probe except one equal to zero.  (The motion of the 
probe for the case with more than one non-zero angular momenta will be 
qualitatively the same.)  Introducing the polar coordinates $(x,
\theta_{\parallel})$ in the rotation plane associated with the non-zero 
angular momentum $J_{\parallel}$, one can express the velocity 
of the probe in the ${\bf x}$-direction as $v^2_{\parallel}=\dot{x}^2+
x^2\dot{\theta}^2_{\parallel}$, where the dot denotes differentiation 
with respect to the time coordinate $\tau$.  
Then, the canonical momenta $p_{\parallel}$ and $p_{\perp}$ along the 
$x$ and $y$ direction, the angular momentum $J_{\parallel}$ and the 
energy $E$ of the probe $p$-brane are as follows:
\begin{eqnarray}
p_{\parallel}&=&m_pH^{{4-\Delta_p}\over{\Delta_p}}_p\dot{x},\ \ \ \ 
p_{\perp}=m_pH^{4\over\Delta}H^{{4-\Delta_p}\over{\Delta_p}}_p\dot{y},\ \ \ \ 
J_{\parallel}=m_px^2H^{{4-\Delta_p}\over{\Delta_p}}_p\dot{\theta}_{\parallel},
\cr
E&=&{{m_p}\over 2}H^{{4-\Delta_p}\over{\Delta_p}}_pv^2_{\parallel}+
{{m_p}\over 2}H^{4\over\Delta}H^{{4-\Delta_p}\over{\Delta_p}}_pv^2_{\perp}
\cr
&=&{{p^2_{\parallel}}\over{2m_pH^{{4-\Delta_p}\over{\Delta_p}}_p}}+
{{J^2_{\parallel}}\over{2m_px^2H^{{4-\Delta_p}\over{\Delta_p}}_p}}+
{{p^2_{\perp}}\over{2m_pH^{4\over\Delta}H^{{4-\Delta_p}\over{\Delta_p}}_p}}.
\label{angmoneng}
\end{eqnarray}

Note, when the motion of the probe is confined along the domain wall 
worldvolume direction, i.e. $v_{\perp}=0$, the probe $p$-brane action 
(\ref{spact}) is independent of the harmonic function $H$ for the domain 
wall, and therefore has the same form as that of a probe $p$-brane in 
the background (\ref{pbrnsol}) of the $(D-1)$-dimensional source $p$-brane.  
However, the harmonic function $H_p$ for the source $p$-brane within the 
$D$-dimensional domain wall has non-trivial modification due to the 
presence of the domain wall.  This implies that the probe $p$-brane can 
distinguish between a $(D-1)$-dimensional $p$-brane (without a domain 
wall) and a $p$-brane within a $D$-dimensional domain wall.  Also, 
unfortunate and unexpected situation is that the harmonic function $H_p$ 
for the source $p$-brane in Eq. (\ref{harmfncs2}) has the radial dependence 
of the form $H_p\sim |{\bf x}|^{-(D-p-8/3)}$ when restricted to the domain 
wall worldvolume directions (i.e., $y=$constant), rather than the usual 
$\sim |{\bf x}|^{-(D-p-4)}$ dependence of a typical $p$-brane in $D-1$ 
dimensions.  Note, the latter dependence was expected because domain walls 
with suitable forms of warp factors are shown to effectively compactify 
the $D$-dimensional gravity to the $(D-1)$-dimensional one through 
gravitational trapping \cite{rs2}, even if the space along the transverse 
direction of the domain walls is non-compact.  

On the other hand, recently it is observed that unlike the case of the KK 
modes of the graviton, the KK zero mode of a massless $U(1)$ gauge boson is 
not localized on the lower-dimensional hypersurface of a domain wall but 
rather spreads over the extra dimensions \cite{pom} and the massive KK modes 
couple to fields on the boundary (at the wall) a lot more strongly than 
the zero mode \cite{dhr,pom}.   (We also expect that such exotic properties 
of the KK modes also hold for the generalization of a $U(1)$ gauge field, 
i.e. a $(p+1)$-form potential, which a $p$-brane couples to.)  Note, in 
such works, it is assumed that the $U(1)$ gauge field does not modify 
spacetime, ignoring the backreaction of the $U(1)$ field to the domain wall 
spacetime and therefore the domain wall spacetime providing with the static 
fixed background in which the $U(1)$ field lives.  However, when the $U(1)$ 
charge becomes large enough for its gravitational backreaction to the domain 
wall spacetime to be non-negligible, such unusual properties of the $U(1)$ 
gauge field KK spectrum will manifest in the Reissner-Nordstrom black hole 
(which is just a spherically symmetric gravitating object with non-zero 
$U(1)$ charge) in one lower dimensions.  We believe that this is the main 
reason for the above mentioned unusual dependence of the $p$-brane harmonic 
function on the radial coordinate.  So, the solution (\ref{pbrdwsol}) may 
correctly describe a $p$-brane in one lower dimensions whose 
higher-dimensional origin is a $p$-brane living inside of the domain wall 
worldvolume, and the existence of such $p$-branes with such unusual radial 
dependence of fields in our world may provide with an evidence that our 
world is embedded inside of the domain wall.  

Therefore, a charged $p$-brane in a $(D-1)$-dimensional world (embedded 
in a $D$-dimensional domain wall) with the expected radial dependence might 
have to be regarded as a $(p+1)$-brane in $D$ dimensions where one of its 
longitudinal directions is along the transverse direction of the domain 
wall.  (So, in particular, a massless $U(1)$ field in the brane world might 
have to be regarded as a two-form potential in one higher dimensions.)  
The configuration is given in the following table.
\begin{center}
\begin{tabular}{|l||c|c|c|c|} \hline
{} \ & \ $t$ \ & \ ${\bf w}$ \ & \ ${\bf x}$ \ & \ $y$ 
\\ \hline\hline
brane \ & \ $\bullet$ \ & \ $\bullet$ \ & \ {} \ & \  $\bullet$ 
\\ \hline
domain wall \ & \ $\bullet$ \ & \ $\bullet$ \ & \ $\bullet$ \ & \ {}  
\\ \hline
\end{tabular}
\end{center}
Here, $t$ is the time coordinate, ${\bf w}=(w_1,...,w_p)$ and $y$ are 
the longitudinal coordinates of the $(p+1)$-brane, and ${\bf w}$ and 
${\bf x}=(x_1,...,x_{D-p-2})$ are the longitudinal coordinates of the 
domain wall.  This point of view is also taken in Ref. \cite{hwa}, where 
it is argued that the Schwarzschild black hole in four-dimensional world 
within a domain wall should be regarded as a black string in five dimensions.  
In fact, since the harmonic function for such $(p+1)$-brane is a harmonic 
function in a $(D-p-2)$-dimensional (conformally) flat space with the 
coordinate ${\bf x}$, the harmonic function for the $p$-brane in one lower 
dimensions will have the expected radial dependence $\sim 
|{\bf x}|^{-(D-p-4)}$.  

Before we will construct such solutions in our future work, it would be 
of interest to check the existence of such solutions and find out about 
the constraint on the dilaton coupling parameters by analyzing the no-force 
condition \cite{nf} on the probe $(p+1)$-brane in the background of the 
source domain wall with the configuration given in the above table.  The 
worldvolume action for the probe $(p+1)$-brane in the background of the 
domain wall is
\begin{equation}
S_{\sigma}=-T_{p+1}\int d^{p+2}e^{-a_{p+1}\phi}\sqrt{-{\rm det}
\left(g^E_{ab}+g^E_{ij}\partial_aX^i\partial_bX^j\right)},
\label{prbpbrndw}
\end{equation}
where $\phi$ and $g^E_{\mu\nu}$ are the fields produced by the source 
domain wall and the Wess-Zumino term does not contribute in this case.  
By expanding the action (\ref{prbpbrndw}) in powers of 
derivatives of $X^i$, one obtains the following effective static 
potential on the probe:
\begin{equation}
V=e^{-a_{p+1}\phi}\sqrt{-g^E_{tt}g^E_{w_1w_1}...g^E_{w_pw_p}g^E_{yy}}=
H^{{4(D+p)-(D-2)^2aa_{p+1}}\over{2(D-2)\Delta}}.
\label{statpotprob}
\end{equation}
So, the force between the probe and the source is balanced, when the 
following constraint on the dilaton coupling parameters is satisfied:
\begin{equation}
aa_{p+1}={{4(D+p)}\over{(D-2)^2}}.
\label{noforce}
\end{equation}
In particular, such configuration is not possible for the non-dilatonic 
domain walls ($a=0$), just like the case of the $p$-brane within the 
domain worldvolume
\footnote{This is one of the reasons why we considered dilatonic 
domain walls in our previous work \cite{youm}.}.  

In the following subsections, we analyze the geodesic motion of the 
probe $p$-brane along the domain wall ($v_{\perp}=0$) and perpendicularly 
to the domain wall ($v_{\parallel}=0$), separately.

\subsection{The motion of the probe along the domain wall}

In this subsection, we study the dynamics of the probe $p$-brane along 
the direction ${\bf x}$ transverse to the source $p$-brane but is 
confined to the worldvolume directions of the domain wall.  In this 
case, $v_{\perp}=0$, i.e., the overall transverse coordinate $z$ is 
constant in time.  We also study the dynamics of a probe $p$-brane moving 
in the background (\ref{pbrnsol}) of the $(D-1)$-dimensional source 
$p$-brane for the purpose of comparison with the former case.    

The dynamic quantities of the probe $p$-brane are
\begin{eqnarray}
p_{\parallel}&=&m_pH^{{4-\Delta_p}\over{\Delta_p}}_p\dot{x},\ \ \ \ \ \ \ \ 
J_{\parallel}=m_px^2H^{{4-\Delta_p}\over{\Delta_p}}_p\dot{\theta}_{\parallel},
\cr
E&=&{{m_p}\over 2}H^{{4-\Delta_p}\over{\Delta_p}}_pv^2_{\parallel}=
{{p^2_{\parallel}}\over{2m_pH^{{4-\Delta_p}\over{\Delta_p}}_p}}+
{{J^2_{\parallel}}\over{2m_px^2H^{{4-\Delta_p}\over{\Delta_p}}_p}},
\label{angengreltr}
\end{eqnarray}
where the source $p$-brane harmonic function is $H_p\approx 
Q_p/\left[x^2+{4\over{9Q^2}}(1+Q|y|)^3\right]^{{3(D-p)-8}\over 6}$ 
for the former case and $H_p\approx Q_p/x^{D-p-4}$ for the latter 
case.  For the incoming probe $p$-brane with the asymptotic velocity $v$ 
and the impact parameter $b$, the asymptotic values (at infinite distance 
from the wall) of the dynamic quantities are $E=m_pv^2/2$, $p_{\parallel}
=m_pv$ and $J_{\parallel}=bm_pv$.  The geodesic motion of such probe 
$p$-brane is then described by the following probe velocity $\dot{x}$ 
along the radial direction $x$, obtained by solving (\ref{angengreltr}):
\begin{equation}
\left|{{dx}\over{d\tau}}\right|=H^{{\Delta_p-4}\over{\Delta_p}}_p
\sqrt{2m_pEH^{{4-\Delta_p}\over{\Delta_p}}_p-{{J^2_{\parallel}}
\over{m^2_px^2}}}=vH^{{\Delta_p-4}\over{\Delta_p}}_p
\sqrt{H^{{4-\Delta_p}\over{\Delta_p}}_p-{{b^2}\over{x^2}}}.
\label{probveldw}
\end{equation}

First, we consider the motion in the background (\ref{pbrnsol}) of the 
$(D-1)$-dimensional source $p$-brane.  There exists the critical 
value of $n\equiv (D-p-4)(4-\Delta_p)/\Delta_p$ below and above which 
the motion of the probe is qualitatively different.  When $n<2$, the 
probe will approach the source with decreasing speed, stop at the 
turning point $x=x_c=(b^2/Q^{(4-\Delta_p)/\Delta_p}_p)^{1/(2-n)}$, 
and re-emerge.  When $n=2$, the probe will approach the source 
monotonically and be ultimately captured by the source, as long 
as $Q^{(4-\Delta_p)/\Delta_p}_p>b^2$.  When $n<2$, the radial motion 
of the probe is the confined within the interval $0\leq x\leq x_c$, 
where $x_c=(b^2/Q^{(4-\Delta_p)/\Delta_p}_p)^{1/(n-2)}$ is the turning 
point, and the probe will be in the end captured by the source.  
On the other hand, for the motion in the background (\ref{pbrdwsol}) 
of the source $p$-brane living in the worldvolume of the source 
domain wall, the probe will never be captured by the source.  When 
$n^{\prime}\equiv{{3(D-p)-8}\over 6}{{4-\Delta_p}\over{\Delta_p}}<1$, 
the probe will approach the source with decreasing speed, stop at 
the turning point and be scattered away.  When $n^{\prime}>1$, the 
radial motion of the probe will be confined to the finite interval 
away from the source, for suitable values of parameters.

\subsection{The motion of the probe perpendicularly to the domain wall}

For the motion of the probe in the direction perpendicular to the 
domain wall, we just consider the background without the source $p$-brane 
for the purpose of seeing whether the domain wall will trap the 
$p$-brane.  In this case, the probe $p$-brane action is given by Eq. 
(\ref{spact}) with $H_p=1$ and $v_{\parallel}=0$.  Therefore, the energy 
$E$ of the probe $p$-brane is
\begin{equation}
E={{m_p}\over{1-H^{4\over\Delta}v^2_{\perp}}},
\label{potdwperp}
\end{equation}
from which we can obtain the following speed of the probe in the 
direction perpendicular to the wall:
\begin{equation}
|v_{\perp}|={\sqrt{E^2-m^2_p}\over{E(1+Q|y|)^{2\over\Delta}}},
\label{probvel}
\end{equation}
and the effective potential on the probe:
\begin{equation}
W(y)=1-{{m_p(E^2-m^2_p)}\over{2E^2(1+Q|y|)^{4\over\Delta}}}.
\label{prbpot}
\end{equation}
So, the velocity of the probe will never be zero at the domain wall 
(located at $y=0$), if $E>m_p$.  When $\Delta>0$, in which case the 
matter is trapped but the gravity is not, the probe $p$-brane will 
approach the domain wall with increasing speed and pass through the 
wall and slow down, approaching zero velocity asymptotically, implying 
that the domain wall attracts the probe $p$-brane.  When $\Delta<0$, 
in which case the matter is not trapped, as the probe approaches the 
domain wall, its speed decreases, and after the probe passes through 
the wall its speed will increase, implying that the domain wall repels 
the probe $p$-brane.

\section{Dynamics of a Test Particle}

In this section, we study the dynamics of an uncharged test particle in 
the background (\ref{pbrdwsol}) of the source $p$-brane living inside of 
the $D$-dimensional domain wall and in the background (\ref{pbrnsol}) of 
the source $p$-brane in $D-1$ dimensions.  Since one of the main purposes 
of such study is to compare the dynamics in these two backgrounds, we 
shall be interested in the motion along the direction ${\bf x}$ transverse 
to the source $p$-brane and restricted to the worldvolume directions 
of the domain wall.  

Generally, the motion of a test particle under the influence of the 
gravitational field $g_{\mu\nu}$ is described by the geodesic equation
\begin{equation} 
{{d^2x^{\mu}}\over{d\lambda^2}}+\Gamma^{\mu}_{\rho\sigma}{{dx^{\rho}}
\over{d\lambda}}{{dx^{\sigma}}\over{d\lambda}}=0,
\label{geodeq}
\end{equation}
where $\Gamma^{\mu}_{\rho\sigma}$ is the Christoffel symbol for the metric 
$g_{\mu\nu}$ and $x^{\mu}(\lambda)$ is the geodesic path of the test 
particle parameterized by an affine parameter $\lambda$.  One can avoid 
having to solve these complicated coupled equations by utilizing the 
symmetry of spacetime in which the test particle moves.  For each Killing 
vector $K^{\mu}$ of spacetime, one can define a constant of motion of the 
test particle by contracting it with the velocity $U^{\mu}=dx^{\mu}/d
\lambda$ of the test particle along the geodesic path $x^{\mu}(\lambda)$.  
For the metrics under consideration in this section, the Killing vectors 
$\partial/\partial t$, $\partial/\partial w_i$ and $\partial/\partial 
\phi_m$ give rise to the following constants of motion for the test particle:
\begin{eqnarray}
E&=&-g^E_{\mu\nu}\left({{\partial}\over{\partial t}}\right)^{\mu}U^{\nu}
=-g^E_{tt}{{dt}\over{d\lambda}},
\cr
p^i&=&g^E_{\mu\nu}\left({{\partial}\over{\partial w_i}}\right)^{\mu}U^{\nu}
=g^E_{ii}{{dw_i}\over{d\lambda}},
\cr
J^m&=&g^E_{\mu\nu}\left({{\partial}\over{\partial \phi_m}}\right)^{\mu}U^{\nu}
=g^E_{\phi_m\phi_m}{{d\phi_m}\over{d\lambda}},
\label{cstmtn}
\end{eqnarray}
where the angular coordinates $0\leq\phi_m<2\pi$ ($m=1,...,[(D-p-2)/2]$) 
are associated with $[(D-p-2)/2]$ rotation planes in $(D-p-2)$-dimensional 
space with the coordinates ${\bf x}$.  In addition, there is another 
constant of motion associated with metric compatibility along the geodesic 
path:
\begin{equation}
\epsilon=-g^{E}_{\mu\nu}{{dx^{\mu}}\over{d\lambda}}{{dx^{\nu}}\over
{d\lambda}},
\label{metcomp}
\end{equation}
where $\epsilon=+1,0,-1$ respectively for a massive particle (i.e. a 
timelike geodesic), a massless particle (i.e. a null geodesic) and a 
spacelike geodesic (even though it does not correspond to the physical 
path of a particle).  

For the test particle moving along the ${\bf x}$-direction, i.e., 
only the ${\bf x}$-component of $U^{\mu}$ is non-zero, the geodesic motion 
is described by the following probe velocity along the radial 
direction $x$  resulting from solving Eqs. (\ref{cstmtn}) and 
(\ref{metcomp}):
\begin{equation}
\left|{{dx}\over{d\lambda}}\right|={1\over\sqrt{g^E_{xx}}}
\sqrt{-{{E^2}\over{g^E_{tt}}}-{{(J^m)^2}\over{g^E_{\phi_m\phi_m}}}
-\epsilon},
\label{xdirecmtn}
\end{equation}
where $\epsilon=1,0$ for a massive and a massless test particle and 
the index $m$ is summed over $m=1,...,[(D-p-2)/2]$.  By plugging the 
explicit expressions for the background metrics of the source branes 
into the above general expression, we obtain
\begin{equation}
\left|{{dx}\over{d\lambda}}\right|=H^{-{{2(p+1)}\over{(D-3)\Delta_p}}}_p
\sqrt{E^2H^{{4(D-p-4)}\over{(D-3)\Delta_p}}_p-{{{\cal J}^2}\over
{x^2H^{{4(p+1)}\over{(D-3)\Delta_p}}_p}}-\epsilon},
\label{xdirecmtn1}
\end{equation}
for the $(D-1)$-dimensional source $p$-brane with $H_p$ given 
in Eq. (\ref{harmdeldefs}), and 
\begin{equation}
\left|{{dx}\over{d\lambda}}\right|=H^{-{2\over{(D-2)\Delta}}}
H^{-{{2(p+1)}\over{(D-2)\Delta_p}}}_p\sqrt{H^{-{4\over{(D-2)\Delta}}}
\left(E^2H^{{4(D-p-3)}\over{(D-2)\Delta_p}}_p-{{{\cal J}^2}\over{x^2
H^{{4(p+1)}\over{(D-2)\Delta_p}}_p}}\right)-\epsilon},
\label{xdirecmtn2}
\end{equation}
for the source $p$-brane in the $D$-dimensional domain wall with 
$H_p$ given in Eq. (\ref{harmfncs2}).  Note, the domain wall harmonic 
function $H$ is just a constant since we consider the motion along the 
longitudinal directions of the domain wall.  Here, ${\cal J}$ is defined 
in terms of the conserved angular momenta $J^m$ of the test particle as
\begin{equation}
{\cal J}^2\equiv \sum^{[{{D-p-2}\over 2}]}_{m=1}{{(L^m)^2}\over{\mu^2_m}},
\label{jdef}
\end{equation}
where $\mu_m$ (whose explicit expressions are given in Ref. \cite{youm2})
are the direction cosines specifying the direction of $r$ and are constant 
due to conservation of the direction of angular momentum (therefore 
${\cal J}$ is also constant).  
By inspecting the expressions (\ref{xdirecmtn1}) and (\ref{xdirecmtn2}) 
for the probe velocities, one can see the following properties for 
the test particle's motion.  First of all, the asymptotic velocity of 
the test particle, as it approaches the $p$-brane, is always non-zero 
and finite for the background (\ref{pbrdwsol}) of $p$-brane in the 
$D$-dimensional domain wall, but is either zero or infinite for the 
background (\ref{pbrnsol}) of $(D-1)$-dimensional $p$-brane.  
The massless test particle $(\epsilon=0)$ will always approach the 
$p$-brane.  But for the massive test particle $(\epsilon=1)$, whereas 
the test particle in the background (\ref{pbrnsol}) will always 
approach the $p$-brane, the test particle in the background 
(\ref{pbrdwsol}) can be scattered away at finite distance from the 
$p$-brane for suitable values of the parameters.  

For the completeness, we just write down the equation describing geodesic 
motion of a test particle along the $y$-direction in the background of 
the source $p$-brane living inside of the domain wall, although it seems 
hard to solve the equation to obtain the expression for the geodesic path.
This can be achieved by considering the geodesic equation (\ref{geodeq}) 
and Eq. (\ref{metcomp}) with only $y$-component of $U^{\mu}$ non-zero, 
resulting in the following equation:
\begin{equation}
{d\over{d\lambda}}\left(g^E_{yy}{{dy}\over{d\lambda}}\right)=
-{\epsilon\over{2g^E_{yy}}}{{\partial g^E_{yy}}\over{\partial y}},
\label{ydrcmtn}
\end{equation}
where $g^E_{yy}=H^{{4(D-1)}\over{(D-2)\Delta}}H^{{4(p+1)}\over{(D-2)
\Delta_p}}_p$.

\section{Conclusion}

In this paper, we studied dynamics of probes in the background of extreme 
dilatonic $p$-branes localized within the worldvolume of extreme dilatonic 
domain walls, for the purpose of studying spacetime properties of such branes 
in comparison with that of the extreme dilatonic $p$-branes in one lower 
dimensions.  We found that the probe dynamics in these two gravitating 
backgrounds do not agree.  We speculated that such disagreement is due to the 
unusual properties of $U(1)$ gauge fields and presumably form potentials that 
their zero modes are not confined within the domain wall.  This led to the 
speculation that an ordinary dilatonic $p$-brane, as observed on the 
hypersurface of the domain wall, should rather be regarded as a dilatonic 
$(p+1)$-brane in the bulk of domain wall, where one of the longitudinal 
directions of the brane is along the direction transverse to the domain wall.  
On the other hand, the existence of a solution describing a $p$-brane 
localized within the domain wall worldvolume implies that such a solution 
may have some physical significance in our world, if the RS type model is the 
true description our nature.  We also discussed the possible 
higher-dimensional embeddings of extreme dilatonic domain wall solutions as 
(intersecting) branes in string theories.  We found that the dilatonic domain 
wall solutions that can be used for the RS type scenario can be obtained 
through the Freund-Rubin compactification of the (intersecting) branes on 
sphere(s).  It seems that dilatonic domain walls are more convenient and 
advantageous than the non-dilatonic domain wall of the original RS model 
\cite{rs1,rs2,rs3} for realizing the RS type scenario, because ($i$) the bulk 
background of the non-dilatonic domain wall does not allow charged branes 
and ($ii$) the most of domain walls obtained by compactifying (intersecting) 
branes in string theories are dilatonic.


\begin{thebibliography} {99}
\small
\parskip=0pt plus 2pt

\bibitem{ncm1} D.W. Joseph, ``Coordinate covariance and the particle 
spectrum,'' Phys. Rev. {\bf 126} (1962) 319.

\bibitem{ncmn}  K. Akama, ``Pregeometry. (Talk),'' {\it In *Nara 1982, 
Proceedings, Gauge Theory and Gravitation*, 267-271}, hep-th/0001113.

\bibitem{ncm2} V.A. Rubakov and M.E. Shaposhnikov, ``Do we live inside a 
domain wall?,'' Phys. Lett. {\bf B125} (1983) 136.

\bibitem{ncm3} V.A. Rubakov and M.E. Shaposhnikov, ``Extra space-time 
dimensions: towards a solution to the cosmological constant problem,'' 
Phys. Lett. {\bf B125} (1983) 139.

\bibitem{ncm4} M. Visser, ``An exotic class of Kaluza-Klein models,'' 
Phys. Lett. {\bf B159} (1985) 22, hep-th/9910093.

\bibitem{ncm5} E.J. Squires, ``Dimensional reduction caused by a cosmological 
constant,'' Phys. Lett. {\bf B167} (1986) 286.

\bibitem{ncm6} P. Laguna-Castillo and R.A. Matzner, ``Surfaces of 
discontinuity in five-dimensional Kaluza-Klein cosmologies,'' 
Nucl. Phys. {\bf B282} (1987) 542.

\bibitem{ncm7} G.W. Gibbons and D.L. Wiltshire, ``Space-time as a membrane 
in higher dimensions,'' Nucl. Phys. {\bf B287} (1987) 717.

\bibitem{rs1} L. Randall and R. Sundrum, ``A large mass hierarchy from a 
small extra dimension,'' Phys. Rev. Lett. {\bf 83} (1999) 3370, 
hep-ph/9905221.

\bibitem{rs2} L. Randall and R. Sundrum, ``An alternative to 
compactification,'' hep-th/9906064.

\bibitem{rs3} J. Lykken and L. Randall, ``The shape of gravity,'' 
hep-th/9908076.

\bibitem{youm} D. Youm, ``Solitons in brane worlds,'' hep-th/9911218.


\bibitem{sg1} K. Behrndt and M. Cveti\v c, ``Supersymmetric domain wall 
world from $D=5$ simple gauged supergravity,'' hep-th/9909058.

\bibitem{sg2} K. Skenderis and P.K. Townsend, ``Gravitational stability 
and renormalization-group flow,'' Phys. Lett. {\bf B468} (1999) 46, 
hep-th/9909070.

\bibitem{sg3} A. Chamblin and G.W. Gibbons, ``Supergravity on the brane,''
hep-th/9909130.

\bibitem{sg4} R. Kallosh, A. Linde and M. Shmakova, ``Supersymmetric 
multiple basin attractors,'' JHEP {\bf 11} (1999) 010, hep-th/9910021.

\bibitem{hwa} A. Chamblin, S.W. Hawking and H.S. Reall, ``Brane-world black 
holes,'' hep-th/9909205.

\bibitem{hm1} R. Emparan, G.T. Horowitz and R.C. Myers, ``Exact description 
of black holes on branes,'' hep-th/9911043.

\bibitem{gs} J. Garriga and M. Sasaki, ``Brane-world creation and black 
holes,'' hep-th/9912118.

\bibitem{hm2} R. Emparan, G.T. Horowitz and R.C. Myers, ``Exact description 
of black holes on branes II: comparison with BTZ black holes and black 
strings,'' hep-th/9912135.

\bibitem{lsol} D. Youm, ``Partially localized intersecting BPS branes,'' 
hep-th/9902208.

\bibitem{ir} R. Argurio, F. Englert and L. Houart, ``Intersection rules for 
$p$-branes,'' Phys. Lett. {\bf B398} (1997) 61, hep-th/9701042.

\bibitem{nf} A.A. Tseytlin, ```No-force' condition and BPS combinations of 
$p$-branes in eleven and ten dimensions,'' Nucl. Phys. {\bf B487} (1997) 141, 
hep-th/9609212.

\bibitem{dw} E. Bergshoeff, M. de Roo, M.B. Green, G. Papadopoulos and 
P.K. Townsend, ``Duality of Type II 7-branes and 8-branes,'' Nucl. Phys. 
{\bf B470} (1996) 113, hep-th/9601150.

\bibitem{pop1} P.M. Cowdall, H. Lu, C.N. Pope, K.S. Stelle and P.K. Townsend,
``Domain walls in massive supergravities,'' Nucl. Phys. {\bf B486} (1997) 49, 
hep-th/9608173.

\bibitem{pop2} H. Lu and C.N. Pope, ``Domain walls from M-branes,'' 
Mod. Phys. Lett. {\bf A12} (1997) 1087, hep-th/9611079.

\bibitem{pop3} I.V. Lavrinenko, H. Lu and C.N. Pope, ``From topology to 
generalized dimensional reduction,'' Nucl. Phys. {\bf B492} (1997) 278, 
hep-th/9611134.

\bibitem{ss} J. Scherk and J.H. Schwarz, ``Spontaneous breaking of 
supersymmetry through dimensional reduction,'' Phys. Lett. {\bf 82B} 
(1979) 60.

\bibitem{fr} P.G. Freund and M.A. Rubin, ``Dynamics of dimensional 
reduction,'' Phys. Lett. {\bf B97} (1980) 233.

\bibitem{gt} G.W. Gibbons and P.K. Townsend, ``Vacuum interpolation in 
supergravity via super $p$-branes,'' Phys. Rev. Lett. {\bf 71} (1993) 3754, 
hep-th/9307049.

\bibitem{ct} P.M. Cowdall and P.K. Townsend, ``Gauged supergravity vacua 
from intersecting branes,'' Phys. Lett. {\bf B429} (1998) 281, 
hep-th/9801165, and references therein.

\bibitem{lps} H. Lu, C.N. Pope, E. Sezgin and K.S. Stelle, ``Dilatonic 
$p$-brane solitons,'' Phys. Lett. {\bf B371} (1996) 46, 
hep-th/9511203.

\bibitem{lpt} H. Lu, C.N. Pope and P.K. Townsend, ``Domain walls from 
anti-de Sitter spacetime,'' Phys. Lett. {\bf B391} (1997) 39, 
hep-th/9607164.

\bibitem{town} H.J. Boonstra, K. Skenderis and P.K. Townsend,
``The domain wall/QFT correspondence,'' JHEP {\bf 01} (1999) 003, 
hep-th/9807137.

\bibitem{bbh} K. Behrndt, E. Bergshoeff, R. Halbersma and J.P. van der Schaar,
``On domain-wall/QFT dualities in various dimensions,'' Class. Quant. Grav. 
{\bf 16} (1999) 3517, hep-th/9907006.

\bibitem{pn} P.K. Townsend and P. van Nieuwenhuizen, 
``Gauged seven-dimensional supergravity,'' Phys. Lett. {\bf B125} (1983) 41.

\bibitem{at} M.A. Awada and P.K. Townsend, ``Gauged $N=4$ $D=6$ 
Maxwell-Einstein supergravity and `antisymmetric tensor Chern-Simons' 
forms,'' Phys. Rev. {\bf D33} (1985) 1557.

\bibitem{rom} L.J. Romans, ``The $F(4)$ gauged supergravity in six 
dimensions,'' Nucl. Phys. {\bf B269} (1986) 691.

\bibitem{gst} M. G\"unaydin, G. Sierra and P.K. Townsend, ``Vanishing 
potentials in gauged $N=2$ supergravity: an application of Jordan algebras,'' 
Phys. Lett. {\bf B144} (1984) 41.

\bibitem{fs} D.Z. Freedman and J.H. Schwarz, ``$N=4$ supergravity theory with 
local $SU(2) \times SU(2)$ invariance,'' Nucl. Phys. {\bf B137} (1978) 333.

\bibitem{dps} M. Douglas, J. Polchinski and A. Strominger, ``Probing 
five-dimensional black holes with D-branes,'' JHEP {\bf 9712} (1997) 003, 
hep-th/9703031.

\bibitem{lt} H. Liu and A.A. Tseytlin, ``Statistical mechanics of D0-branes 
and black hole thermodynamics,'' JHEP {\bf 9801} (1998) 010, hep-th/9712063.

\bibitem{kir1} E. Kiritsis and T.R. Taylor, ``Thermodynamics of D-brane 
probes,'' hep-th/9906048.

\bibitem{kir2} E. Kiritsis, ``Supergravity, D-brane probes and thermal super 
Yang-Mills:  A comparison,'' JHEP {\bf 9910} (1999) 010, hep-th/9906206.

\bibitem{kir3} A. Kehagias and E. Kiritsis, ``Mirage cosmology,''
JHEP {\bf 9911} (1999) 022, hep-th/9910174.

\bibitem{pom} A. Pomarol, ``Gauge bosons in a five-dimensional theory with 
localized gravity,'' hep-ph/9911294.

\bibitem{dhr} H. Davoudiasl, J.L. Hewett and T.G. Rizzo, 
``Bulk gauge fields in the Randall-Sundrum model,'' hep-ph/9911262.

\bibitem{youm2} M. Cveti\v c and D. Youm, ``Near-BPS-saturated rotating 
electrically charged black holes as string states,'' Nucl. Phys. {\bf B477} 
(1996) 449, hep-th/9605051.

\end{thebibliography}
\end{document}